\documentclass[
prc,%
10pt,%
final,%
notitlepage,%
oneside,%
twocolumn,%
nobibnotes,%
nofootinbib,
superscriptaddress,%
floatfix,%
floatfix,%
showkeys,%
showpacs]%
{revtex4}
\usepackage{color}
\usepackage{amsfonts}
\usepackage{amsbsy}
\usepackage{mathrsfs}
\usepackage{graphicx}
\def\lsim{\mathrel{\rlap{
\lower4pt\hbox{\hskip-3pt$\sim$}}
    \raise1pt\hbox{$<$}}}     
\def\gsim{\mathrel{\rlap{
\lower4pt\hbox{\hskip-3pt$\sim$}}
    \raise1pt\hbox{$>$}}}     
\def\scr#1{\mbox{\scriptsize #1}}
\begin{document}
\title{
High baryon and energy densities achievable in heavy-ion
collisions at $\sqrt{s_{NN}}= 39$ GeV} 
%
\author{Yu. B. Ivanov}\thanks{e-mail: Y.Ivanov@gsi.de}
\affiliation{National Research Centre "Kurchatov Institute", Moscow 123182, Russia} 
\affiliation{National Research Nuclear University "MEPhI",  
Moscow 115409, Russia}
\affiliation{Bogoliubov Laboratory of Theoretical Physics, JINR, Dubna 141980, Russia}
\author{A. A. Soldatov}
\affiliation{National Research Nuclear University "MEPhI",  
Moscow 115409, Russia}
\begin{abstract}
Baryon and energy densities, which are reached in central Au+Au collisions at collision energy 
of $\sqrt{s_{NN}}= 39$ GeV, are estimated within the model of three-fluid dynamics. 
It is shown that the initial thermalized mean proper baryon and energy densities  
in a sizable central region approximately are
$n_B/n_0 \approx$ 10 and $\varepsilon\approx$ 40 GeV/fm$^3$, respectively. 
The study indicates that the deconfinement transition at the stage of interpenetration of colliding nuclei 
makes the system quite opaque. The final fragmentation regions in these collisions 
are formed not only by primordial fragmentation fireballs, i.e. the baryon-rich matter  passed through 
the interaction region (containing approximately 30\% of the total baryon charge), but also by the 
baryon-rich regions of the central fireball pushed out to peripheral rapidities 
by the subsequent almost one-dimensional expansion of the central fireball along the beam direction.
\end{abstract}
\maketitle


At ultra-relativistic  energies, colliding nuclei pass through
each other, compressing and depositing energy in each other,  
rather than mutually stopping, as at lower energies. 
The net-baryon charge remains concentrated in the fragmentation regions
that are well separated in the configuration and momentum space from the mid-rapidity fireball.
Properties of these baryon-rich fragmentation regions (i.e. the baryonic fireballs)
produced in central  
heavy-ion collisions were discussed long ago 
\cite{Anishetty:1980zp,Csernai:1984qh,Gyulassy:1986fk,Frankfurt:2002js,Mishustin:2001ib}. 
Recent proposal \cite{Brodsky:2012vg}
to perform experiments at the Large Hadron Collider (LHC) at CERN in the
fixed-target mode  (AFTER@LHC experiment) revived interest to 
the fragmentation regions. 
This experiment would provide an opportunity
to carry out precision measurements in the kinematical 
region of the target fragmentation region. 
If the LHC operates in a fixed-target mode at a beam
energy of 2.76 GeV per nucleon, 
this is equivalent to $\sqrt{s_{NN}}=$ 72 GeV in terms of the center-of-mass
energy. This energy is only slightly above the range of  
the Beam Energy Scan (BES) program
at the BNL Relativistic Heavy Ion Collider (RHIC).  

Recently theoretical considerations 
on the internal properties of baryonic fireballs
were updated in Refs.
\cite{Li:2016wzh} based on the McLerran-Venugopalan model \cite{McLerran:1993ka}.
It was argued \cite{Li:2016wzh} that in central Au+Au collisions 
at the top RHIC energy, 
high baryon densities (an order of magnitude greater than the normal nuclear one) 
over a large volume are achieved in fireballs outside the central rapidity region. 
This is in contrast to almost net-baryon-free matter produced in the midrapidity region.  
However, the LHC energy in the fixed-target mode,  
 provides only
$\sqrt{s_{NN}}=$ 72 GeV, which is already near the lower limit of applicability
of the McLerran-Venugopalan model \cite{Li:2016wzh}. Therefore, phenomenological approaches are required 
for estimation of the baryon and energy densities 
reached in the fragmentation regions at these energies.

In the present paper 
we estimate the baryon and energy densities 
reached in central Au+Au collisions
within the model of the three-fluid dynamics
(3FD) \cite{3FD,Ivanov:2013wha}.  
The estimation is done for  
the highest collision energy of 39 GeV accessible for the 3FD simulations. 
This energy is certainly lower than the top LHC energy in the fixed-target mode, 
however the main features of the fragmentation regions are expected to be similar to 
those at 72 GeV. 
The 3FD model is quite successful in reproducing 
the major part of the 
observables in the midrapidity region at the BES RHIC energies 
\cite{Ivanov:2013wha,Ivanov:2012bh,Ivanov:2013yqa,Ivanov:2013yla,Ivanov:2016sqy,Ivanov:2014zqa,3FD-bulk-RHIC}.
Therefore, the 3FD predictions for the fragmentation regions may be of interest.


Unlike conventional hydrodynamics, where local
instantaneous stopping of projectile and target matter is
assumed, a specific feature of the 3FD description \cite{3FD} 
is a finite stopping power resulting in a counterstreaming
regime of leading baryon-rich matter. This generally
nonequilibrium regime of the baryon-rich matter
is modeled by two interpenetrating baryon-rich fluids 
initially associated with constituent nucleons of the projectile
(p) and target (t) nuclei. In addition, newly produced particles,
populating the midrapidity region, are associated with a fireball
(f) fluid.
Each of these fluids is governed by conventional hydrodynamic equations 
coupled by friction terms in the right-hand sides of the Euler equations. 
{
These friction terms describe energy--momentum loss of the 
baryon-rich fluids. A part of this
loss is transformed into thermal excitation of these fluids, while another part 
gives rise to particle production into the fireball fluid.
}

Friction forces between fluids are the key constituents of the model that 
determine dynamics of the nuclear collision. 
The friction forces in the hadronic phase were estimated in Ref. \cite{Sat90}. 
Precisely these friction forces are used in the simulations 
for the hadronic phase. 
There are no theoretical estimates of
the friction in the quark-gluon phase (QGP) so far.
Therefore, the friction in the QGP is purely phenomenological.
It was fitted to reproduce the baryon
stopping at high collision energies within the deconfinement
scenarios as it is described in  Ref. \cite{Ivanov:2013wha} in detail.

The physical input of the present 3FD calculations is described in
Ref.~\cite{Ivanov:2013wha}. 
The simulations in 
\cite{Ivanov:2013wha,Ivanov:2012bh,Ivanov:2013yqa,Ivanov:2013yla,Ivanov:2016sqy,Ivanov:2014zqa,3FD-bulk-RHIC} 
were performed with different 
equations of state (EoS's)---a purely hadronic EoS \cite{gasEOS}  
and two versions of the EoS involving the   deconfinement
 transition \cite{Toneev06}, i.e. a first-order phase transition  
and a smooth crossover one. In the present paper we demonstrate results with 
only these deconfinement
EoS's as the most successful in reproduction of various 
observables at high collision energies:  
the baryon stopping \cite{Ivanov:2013wha,Ivanov:2012bh}, 
yields of various hadrons \cite{Ivanov:2013yqa}, their mean transverse masses \cite{Ivanov:2013yla,Ivanov:2016sqy}, 
the elliptic flow \cite{Ivanov:2014zqa}, etc. 
{A detailed comparison with the recent STAR data on bulk observables \cite{Adamczyk:2017iwn}
is presented in Ref. \cite{3FD-bulk-RHIC}.
} 
%
Due to numerical reasons \cite{3FD}, 39 GeV is the highest energy attainable for computations 
within the 3FD model.

For the discussion below we need to introduce some quantities. 
Within the 3FD model the system is characterized by three hydrodynamical velocities,  
$u_{\alpha}^{\mu}$ with $\alpha=$ p, t and f, attributed to these fluids. 
The interpenetration of the p and t  fluids takes place 
only at the initial stage of the nuclear collision. At later stages either 
a complete mutual stopping occurs and these fluids get unified or these fluids
become spatially separated.   Therefore, 
we define a collective 4-velocity of the baryon-rich matter associating it 
with the total baryon current    $u^{\mu}_B =  J_{B}^{\mu}/|J_{B}|$, 
%
%
where $J_{B}^{\mu} = n_{\scr p}u_{\scr p}^{\mu}+n_{\scr t}u_{\scr t}^{\mu}$ is the baryon
current defined in terms of proper baryon densities $n_{\alpha}$ of these fluids and    
hydrodynamic 4-velocities $u_{\alpha}^{\mu}$, and 
   \begin{eqnarray}
   \label{nb-prop}
   |J_{B}|= \left(J_{B}^{\mu} J_{B\mu}\right)^{1/2}\equiv n_B
   \end{eqnarray}
is the proper (i.e. in the local rest frame) baryon density of the p and t  fluids. 
In particular, this proper baryon density allows us to construct a simple fluid 
unification measure 
   \begin{eqnarray}
   \label{unification}
   1-\frac{n_{\scr p}+n_{\scr t}}{n_B}
   \end{eqnarray}
which is zero, when the p and t  fluids are mutually stopped and unified, and  
has a positive value increasing  with rise of
the relative velocity of the p and t  fluids.

The total proper energy density of all three fluids in the local rest frame,  
where the composed matter is at rest, is defined as follows
\begin{eqnarray}
\label{eps_tot}
\varepsilon = u_\mu T^{\mu\nu} u_\nu. 
\end{eqnarray}
This proper energy density is
defined in terms of the total energy--momentum tensor
$T^{\mu\nu} \equiv
T^{\mu\nu}_{\scr p} + T^{\mu\nu}_{\scr t} + T^{\mu\nu}_{\scr f}$
%
%
being the sum of conventional hydrodynamical energy--momentum tensors of separate fluids, and
the total collective 4-velocity of the matter
\begin{eqnarray}
\label{u-tot}
u^\mu = u_\nu T^{\mu\nu}/(u_\lambda T^{\lambda\nu} u_\nu). 
\end{eqnarray}
Note that definition (\ref{u-tot}) is, in fact, an equation
determining $u^\mu$. In general, this $u^\mu$ does not coincide with 
4-velocities of separate fluids. 
This definition is in the spirit of the
Landau--Lifshitz approach to viscous relativistic hydrodynamics.


%
\begin{figure}[!htb]
\includegraphics[width=8.6cm]{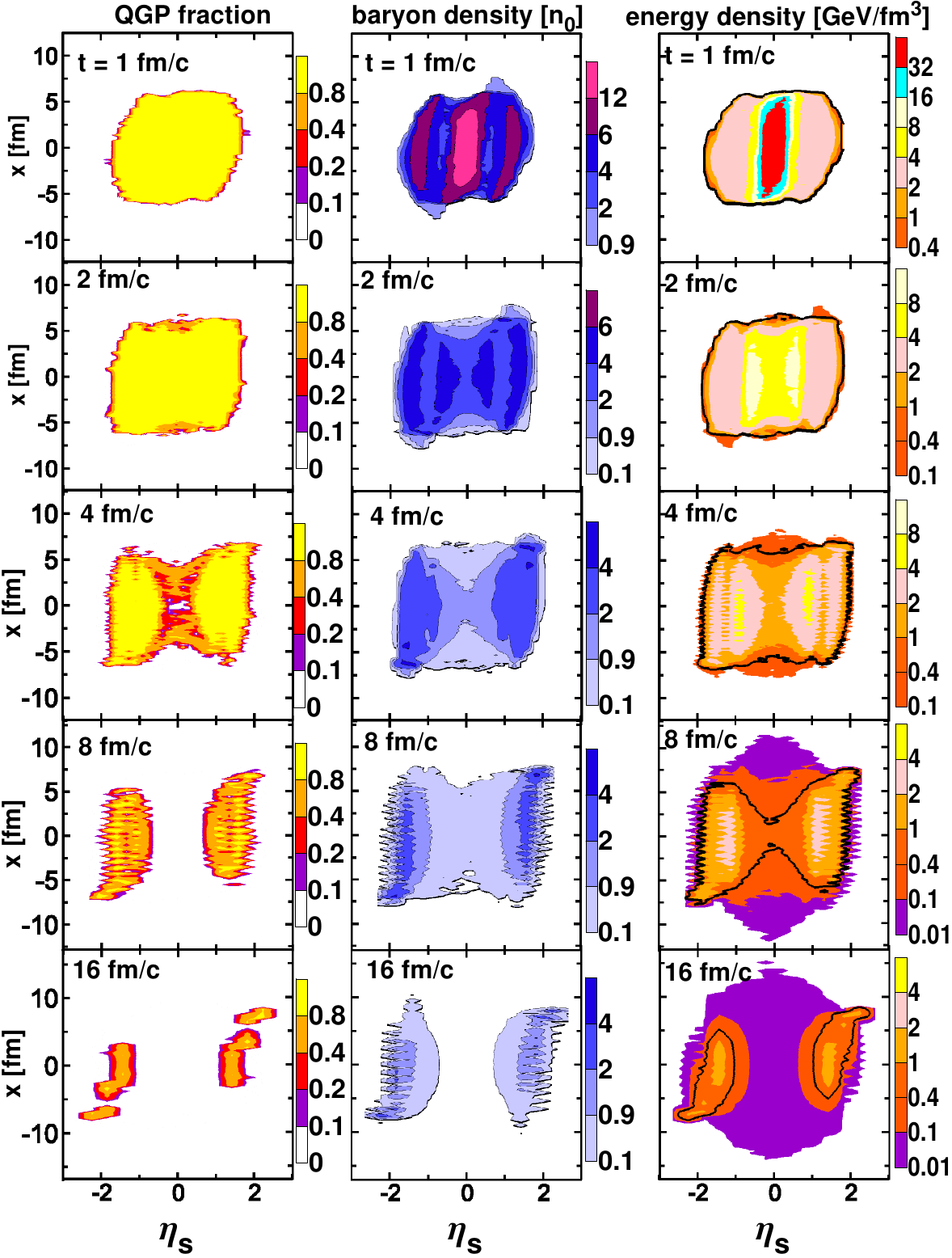}
 \caption{(Color online)
QGP fraction (left column), the  proper baryon density in units of 
the normal nuclear density, $n_0=0.15$ 1/fm$^3$, see Eq. (\ref{nb-prop}) (middle column),   
and proper energy density, see Eq. (\ref{eps_tot}) (right column),  
in the reaction plain ($x\eta_s$) at various time instants (in the c.m. frame) 
in the central ($b=$ 2 fm) Au+Au collision at $\sqrt{s_{NN}}=$ 39 GeV. 
$\eta_s$ is the space-time rapidity along the beam direction. 
Calculations are done with the first-order-transition EoS. 
The bold contours in panels of the right column 
display the borders between the frozen-out and still hydrodynamically
evolving matter.  
}
\label{fig2}
\end{figure}
\begin{figure}[!htb]
\includegraphics[width=8.6cm]{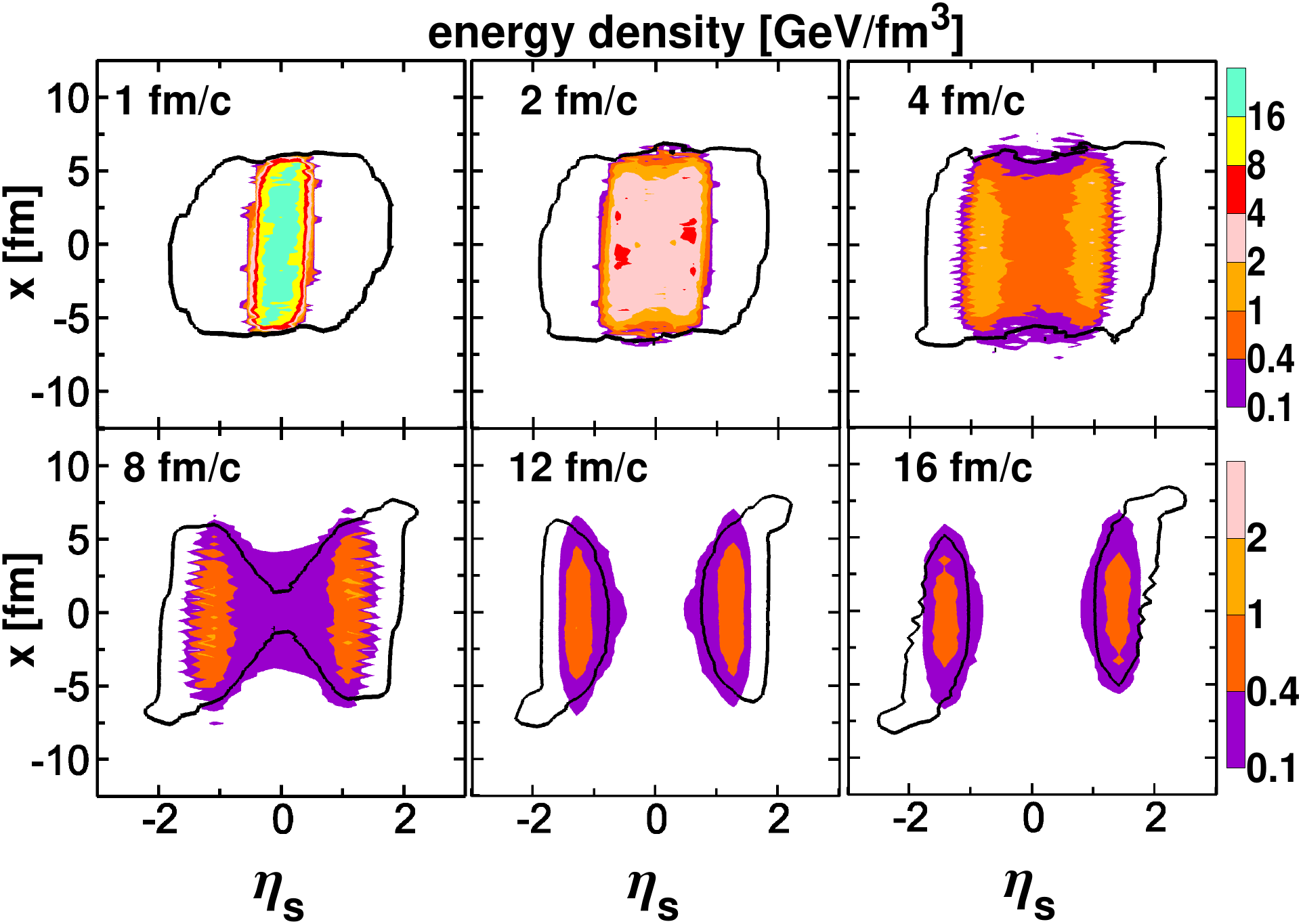}
 \caption{(Color online)
Proper energy density of the baryon-free (f) fluid  
in the reaction plain ($xz$) at various time instants 
in the central ($b=$ 2 fm) Au+Au collision at $\sqrt{s_{NN}}=$ 39 GeV. 
Calculations are done with the first-order-transition EoS. 
The bold contours  
display the borders between the frozen-out and still hydrodynamically
evolving matter.  
}
\label{fig2a}
\end{figure}

Figure \ref{fig2} presents the time evolution of the QGP fraction and    
the proper baryon and energy densities, Eqs. (\ref{nb-prop}) and (\ref{eps_tot}), 
respectively, 
{
in the reaction plain ($x\eta_s$) of
central Au+Au collision at $\sqrt{s_{NN}}=$ 39 GeV, where 
   \begin{eqnarray}
   \label{eta_s}
   \eta_s = \frac{1}{2} \ln\left(\frac{t+z}{t-z}\right)
   \end{eqnarray}
is the space-time rapidity and $z$ is the coordinate along the beam direction. 
}
The baryon-rich fluids are  
mutually stopped and unified already at $t\gsim 1$ fm/c because the fluid unification measure, 
see Eq.  (\ref{unification}),   
is practically zero (less than 0.02). 
The baryon-fireball relative velocity is small,  
$v_{{\scr f}B} \lsim 0.1$, at $t \geq$ 1 fm/c. This indicates that a system is close 
to the thermal (kinetic) equilibrium. 
{
As the the f-fluid is not that well unified with the combined baryon-rich pt-fluid, 
the evolution of the f-fluid is separately presented in Fig. \ref{fig2a}. 
The pt-fluid entrains the f-fluid. This is the reason for the smallness of $v_{{\scr f}B}$. 
}

As seen from Fig. \ref{fig2}, at $t=$ 1 fm/c
the matter of colliding nuclei has already partially passed though 
the interaction zone (two bumps of baryon density near $\eta_s=\pm$ 1) 
and has been partially stopped in the central region (the central bumps in $n_B$ and $\varepsilon$). 
Thus, the central region and the fragmentation regions have already been formed 
to $t=$ 1 fm/c. The matter in all these regions is in the quark-gluon phase, 
see the QGP fraction in Fig. \ref{fig2}.  
A large fraction of the baryon charge stopped in 
{
the central region ($\approx$ 70\%) 
is in contrast 
to the ultra-relativistic  scenario (at the top RHIC and LHC energies) where the major 
part of the baryon charge is assumed to be located in the fragmentation regions
already at the initial stage. 
The proper baryon and energy densities  in this central region approximately are
$n_B/n_0 \approx$ 10 and $\varepsilon\approx$ 40 GeV/fm$^3$, respectively. 
The present situation is more similar to that at moderate energies, 
as predicted by transport models \cite{Nara:1999dz,Cassing:2009vt,Bass:1998ca,Amelin:1989vp,Lin:2004en}. 
}

%
\begin{figure}[!tbh]
\includegraphics[width=8.5cm]{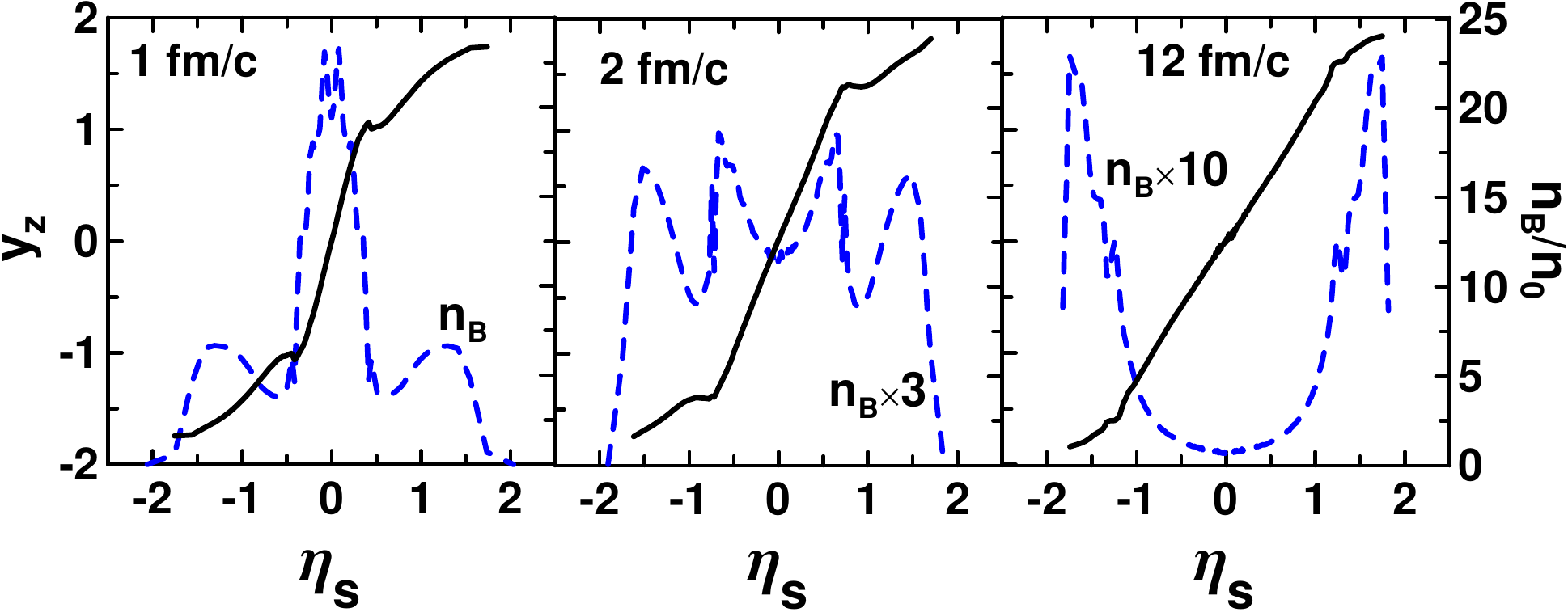}
 \caption{(Color online)
Proper baryon density in units of 
the normal nuclear density
(dashed lines, right scale axis)  
and 
the longitudinal rapidity ($y_z$) 
of the matter (solid lines, left scale axis)
 along the beam axis ($x=y=0$)
at various time instants 
in the central ($b=$ 2 fm) Au+Au collision simulated within 
the first-order-transition scenario. 
}
\label{fig3}
\end{figure}

The fine structure of the evolving system along the beam axis ($\eta_s$, $x=y=0$)
is presented in Fig. \ref{fig3}.
As seen, the central region undergoes a  
rapid, practically self-similar one-dimensional (1D) expansion right after its thermalization. 
This expansion pushes out the outer layers of the central fireball while the 
inner region serves as a driving force. 
The primordial fragmentation fireballs also expand in counter directions to the central one. 
The effect of this counter expansion is seen as wiggles 
in the $\eta_s$ dependence of the longitudinal rapidity, 
   \begin{eqnarray}
   \label{y_z}
   y_z = \frac{1}{2} \ln\left(\frac{1+v_z}{1-v_z}\right),
   \end{eqnarray}
at the borders between the fragmentation and central fireballs 
see $t$ = 1 and 2 fm/c panels in Fig. \ref{fig3}.
Here $v_z$ is the $z$ component of the hydrodynamical 3-velocity [Eq. (\ref{u-tot})].   
The positions of these wiggles 
precisely coincide with the borders between the f-fluid and the primordial fragmentation fireballs, 
see Fig. \ref{fig2a}.

{
In the course of time, this predominantly 1D expansion of the central fireball
further proceeds, see Figs. \ref{fig2} and \ref{fig3}. 
The matter, and in particular the baryon charge, is pushed out to the periphery of this central fireball, 
i.e. closer to the primordial fragmentation regions. 
The primordial fragmentation fireballs join 
with ``central'' contributions to the  
instant $t=4$ fm/c because of their counter expansion, see Fig. \ref{fig2}.   
At t = 12 fm/c only tiny wiggles on the inner slopes of the density peaks and 
the corresponding tiny wiggles in the rapidity profile indicate this joining, 
see Fig. \ref{fig3}. 
Therefore, the final fragmentation regions consist of primordial fragmentation fireballs,  
i.e. the baryon-rich matter  passed through the interaction region, 
and baryon-rich regions of the central fireball pushed out to peripheral rapidities. 
However, full mixing of these ``central'' and primordial fragmentation fireballs   
does not occur---the primordial fragmentation regions do not overlap 
with the f-fluid even at late time instants, as seen from Fig. \ref{fig2a}.  
}

\begin{figure}[!tbh]
\includegraphics[width=7.cm]{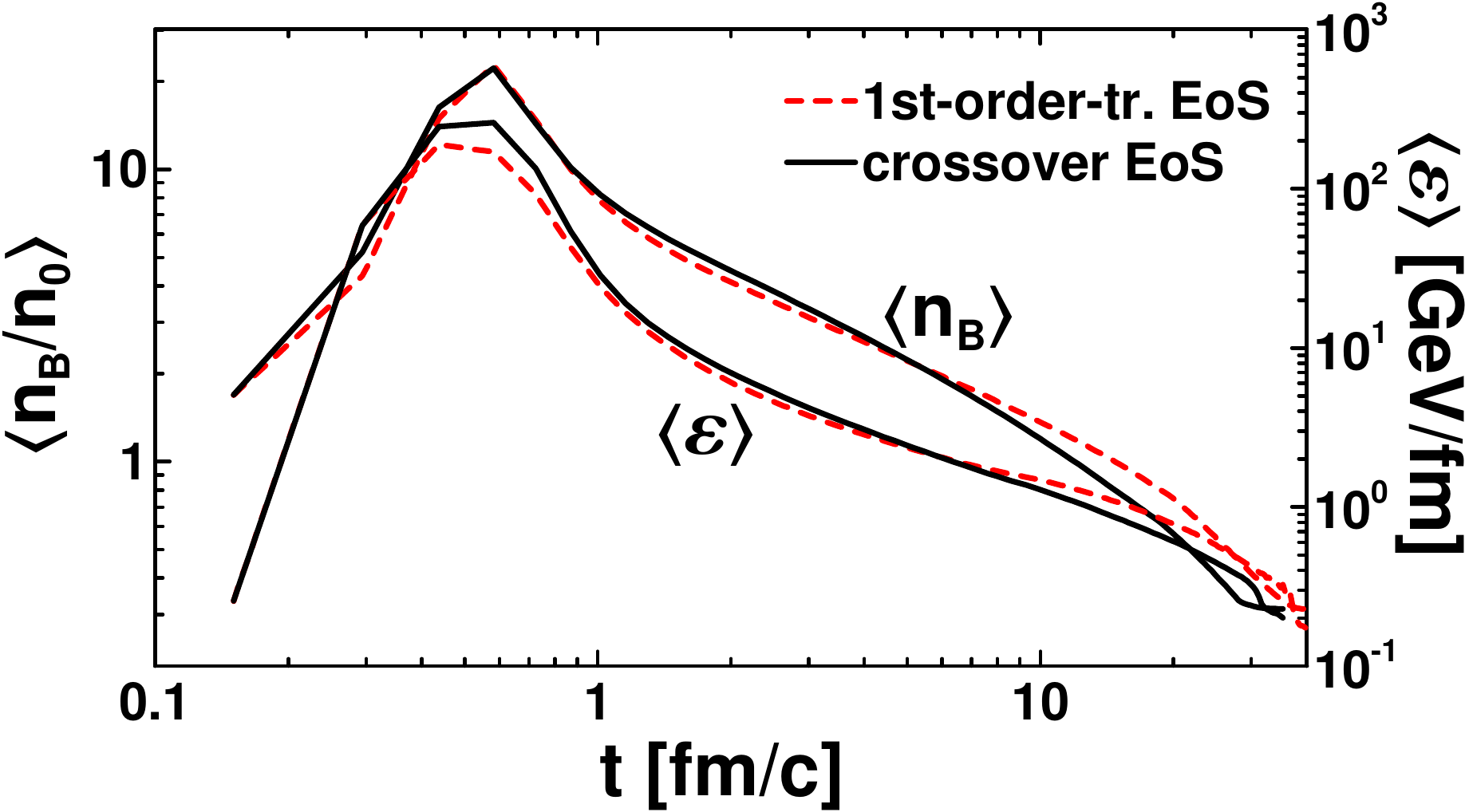}
 \caption{(Color online)
Time evolution of the mean proper baryon (left scale) and energy (right scale) densities
in central 
Au+Au collisions at $\sqrt{s_{NN}}=$ 39 GeV 
in simulations with different EoS's.  
}
\label{fig4}
\end{figure}

At later time $t \geq$ 10 fm/c, see Fig. \ref{fig2}, 
the central part of the system gets frozen out while the fragmentation regions continue 
to evolve being already separated in the configuration space. 
This longer evolution of the fragmentation regions is due to the relativistic time dilation 
caused by their high-speed motion with respect to the central region. 
{
Therefore, their evolution time in the c.m. frame of colliding nuclei
lasts $\approx 40$ fm/c, as seen in Fig. \ref{fig4}. 
}

{
To gain an impression of 
the proper baryon and energy densities attainable in a sizable volume,  
}
we present the evolution of mean proper densities
averaged over the whole volume of a still hydrodynamically evolving system
in Fig. \ref{fig4}. 
[Note that the freeze-out in the 3FD model removes the frozen-out matter from the hydrodynamical 
evolution \cite{Russkikh:2006aa}.] 
%
These values are very similar for the first-order-transition and crossover EoS's. 
Note that this similarity is not due to similarity of these two EoS's. 
This similarity takes place because of the friction forces that were independently fitted 
for each EoS in order to reproduce observables in the midrapidity region.


In conclusion, 
{
it is demonstrated that at the initial thermalized stage of the central Au+Au collision 
only  $\approx$30\% of the
baryon charge is located in fragmentation regions,
while $\approx$70\% -- in the central fireball.
%
The initial thermalized proper baryon and energy densities  approximately are
$n_B/n_0 \approx$ 10 and $\varepsilon\approx$ 40 GeV/fm$^3$, respectively. 
If the calculation is performed with the hadronic friction \cite{Sat90}, 
we obtain a very high transparency of the colliding nuclei and 
at the same time do not reproduce the experimental data \cite{Adamczyk:2017iwn}. 
Therefore, the present results indicate that the transition into the QGP 
at the stage of interpenetration of colliding nuclei 
makes the system more opaque.
Alternatively they may indicate a formation of strong color fields between the leading
partons \cite{Mishustin:2001ib} preceding the QGP production. 
These fields may enhance baryon stopping as compared to its estimate 
based on hadronic cross-sections \cite{Sat90}.

Though these high densities are formed in the central fireball, their 
observable consequences manifest themselves
in the fragmentation regions 
where this dense matter
is pushed out by the subsequent fast 
1D expansion of the central fireball along the beam direction.
Thus, the final fragmentation regions in the central Au+Au collisions 
at $\sqrt{s_{NN}}=$ 39 GeV are formed by
not only primordial fragmentation fireballs,  
i.e. the baryon-rich matter  passed through the interaction region, but also by the 
baryon-rich regions of the central fireball pushed out to peripheral rapidities. 
}
It is expected that the role of this central fireball gradually reduces with the collision energy rise 
and the dense baryon matter becomes predominantly located in the primordial fragmentation fireballs
already at the initial stage of the collision. 


This work was carried out using computing resources of the federal collective usage center «Complex for simulation and data processing for mega-science facilities» at NRC "Kurchatov Institute", http://ckp.nrcki.ru/.
Y.B.I. was supported by the Russian Science
Foundation, Grant No. 17-12-01427.
A.A.S. was partially supported by  the Ministry of Education and Science of the Russian Federation within  
the Academic Excellence Project of 
the NRNU MEPhI under contract 
No. 02.A03.21.0005. 

\end{document}